\documentclass[%
 preprint,
 amsmath,amssymb,
 aps,
]{revtex4-1}

\usepackage{graphicx}
\usepackage{dcolumn}
\usepackage{bm}

\def\lb{\label}

\def\/{\over}

\def\el2{\ell^{\,2}}             \def\1{1\!\!1}

\def\const{\mathop{\mathrm{const}}\nolimits}

\newcommand{\ca}{\begin{cases}}
\newcommand{\ac}{\end{cases}}
\newcommand{\ma}{\begin{pmatrix}}
\newcommand{\am}{\end{pmatrix}}
\renewcommand{\[}{\begin{eqnarray}}
\renewcommand{\]}{\end{eqnarray}}
\def\eq{\begin{equation}}
\def\qe{\end{equation}}

\begin{document}


\title{Quasistatic stopband in the spectrum of one-dimensional piezoelectric
phononic crystal}

\author{A.A. Kutsenko}
\author{A.L. Shuvalov}%
\author{O. Poncelet}

\affiliation{Universit\'{e} de Bordeaux, CNRS, UMR 5295, Institut de
M\'{e}canique et d'Ing\'{e}nierie de Bordeaux,
Talence 33405, France}%

\author{A.~N.~Darinskii}
\affiliation{ Institute of Crystallography RAS, 119333 MOSCOW,
RUSSIA}



\date{\today}

\begin{abstract}
Propagation of a longitudinal wave through the periodic structure
composed of alternating elastic and piezoelectric layers is
considered. The faces of each piezoelectric layer are electroded and
connected via a circuit with the capacity $C$. It is shown that if
$C<0$ then the Floquet-Bloch spectrum $ \omega(K)$ in a certain
range of negative $C$ may possess a quasistatic absolute stopband
starting at $\omega =0$. Other unusual features of the spectrum
occurring at certain fixed values of $C<0$ are the infinite group
velocity of the first branch at the origin point $\omega =0$, $K=0$
and the flat bands $\omega=\const$.
\end{abstract}

\pacs{Valid PACS appear here}
\maketitle


\section{\label{int}Introduction}

A periodic structure composed of alternating elastic and
piezoelectric layers, where the faces of each piezoelectric layer
are electroded and connected by a circuit with the capacity $C$, has
been considered in \cite{1,2}. The dispersion equation for a
bilayered unit cell has been obtained in \cite{1} and generalized
for a multilayered unit cell in \cite{2}. The tunable via $C$
effective constants of the periodic structure were derived in
\cite{2}. It has been noted in \cite{1} that the capacity $C$ of the
external circuit may be negative. The present paper analyzes some
unusual features of the Floquet-Bloch dispersion spectrum occurring
at $C<0.$ In such a case, using the definition of the quasistatic
effective elastic constant $c_{\mathrm{eff}}^{0}$ \cite{2}, it is
shown that the most interesting behavior comes about at low
frequency. This is because the value of $c_{\mathrm{eff}}^{0}$
becomes negative in the certain interval of $C<0$, where
$c_{\mathrm{eff}}^{0}$ varies from $-\infty $ to zero. As a result,
at $c_{\mathrm{eff}}^{0}<0$ there occurs an absolute stopband which
starts from the quasistatic limit $\omega =0$, while the first
dispersion branch starts from nonzero $\omega$. At certain negative
$C$, the group velocity of this branch changes from positive to
negative value. Other peculiar features are infinite group velocity
at the origin $\omega =0,~K=0$ of the first dispersion branch at the
pole $c_{\mathrm{eff}}^{0}\rightarrow +\infty $ and a flat band
$\omega =0$ at $c_{\mathrm{eff}}^{0}=0$.

\section{\label{back}Background}

Consider a periodic structure whose unit cell consists of elastic
layer 1 and piezoelectric layer 2. Suppose that the faces of each
piezoelectric layer are electroded and connected by a circuit with
capacity $C.$ Let a time-harmonic longitudinal wave propagate along
the axis $X_{3}$ orthogonal to the layer interfaces. Denote the
density and elastic constant of the elastic layer by $\rho _{1}$ and
$c_{33}=c_{1},$ and the density and elastic, piezoelectric and
dielectric constants of the piezoelectric layer by $\rho _{2}$,
$c_{33}^{E}\equiv c^{E}$, $c_{33}^{D}\equiv
c^{D}=c^{E}+e^{2}/\varepsilon $, $e_{33}\equiv e$,$\ \varepsilon
_{33}\equiv \varepsilon .$ The layer thicknesses are $d_{1}$ and
$d_{2}$ so that the period is $T=d_{1}+d_{2}.$ The displacement
$u_{3}\equiv u$ and stress$\ \sigma _{33}\equiv \sigma $ at the
period edges $x_{3}=0$ and $x_{3}=T$ are related as follows \cite{2}
\begin{equation}
\begin{pmatrix}
u( T) \\
\sigma( T)%
\end{pmatrix} =\mathbf{m}_2\mathbf{m}_1\begin{pmatrix}
u( 0) \\
\sigma ( 0)%
\end{pmatrix}  \label{9}
\end{equation}%
with
\[
\mathbf{m}_1=\mathbf{m}_{10},\ \ \mathbf{m}_2=\mathbf{m}_{20}+
\frac{1}{\frac{S}{C}-M_{3}}\begin{pmatrix}
M_{1} \\
M_{2}%
\end{pmatrix} \begin{pmatrix}
M_{2} \\
M_{1}%
\end{pmatrix} ^{\top},\notag\\
\mathbf{m}_{j0}=\begin{pmatrix}
\cos k_j d_j & \frac{\sin k_jd_j}{Z_j\omega} \\
-Z_j\omega\sin k_jd_j & \cos k_j d_j
\end{pmatrix},\ \ \ j=1,2.\ \ \ \ \ \label{9a}
\]
where $S$ is the surface area of the electrode, $^{\top}$ means
transposition and
\begin{eqnarray}
M_{1}=\frac{h\sin k_2d_2}{Z_2\omega},\ M_{2}=h(\cos k_2 d_2-1),\notag\\
M_{3}=hM_{1}-\frac{d_2}{\varepsilon},\ \
k_1=\omega\sqrt{\frac{\rho_1}{c_1}},\ \
k_2=\omega\sqrt{\frac{\rho_2}{c_2^{D}}},\notag\\
h=\frac{e}{\varepsilon},\ \ Z_1=\sqrt{\rho_1 c_1},\ \
Z_2=\sqrt{\rho_2 c_2^{D}}.\ \ \ \ \ \label{6}
\end{eqnarray}
In view of (\ref{9}), the dispersion equation of the Floquet-Bloch
spectrum $\omega(K)$ is
\begin{equation}
\cos KT=\frac{1}{2}\mathrm{trace~}(\mathbf{m}_2\mathbf{m}_1),
\label{11}
\end{equation}
where $K$ is the Floquet wavenumber. According to \cite{2}, taking
low-frequency asymptotics yields the quasistatic effective elastic
constant and effective density of the periodic structure in the form
of
\begin{equation}
\frac{T}{c_{\mathrm{eff}}}=\frac{d_{1}}{c_{1}}+d_{2}\left( c_{2}^{E}+%
\frac{e^{2}}{\frac{Cd_{2}}{S}+\varepsilon }\right)^{-1},\ \ \ \
T\rho_{\rm
 eff}=d_1\rho_1+d_2\rho_2.
\label{a9}
\end{equation}
The corresponding quasistatic effective speed is
\[\lb{es}
 \lim_{K\to0}\frac{\omega}{K}=\sqrt{\frac{c_{\rm eff}}{\rho_{\rm
 eff}}}.
\]
Our purpose is to analyze the dispersion spectrum and its
low-frequency limit in the case $C<0$.

\section{\label{examp}Example}

\begin{figure}[h]
\begin{minipage}[h]{0.2\linewidth}
\center{\includegraphics[width=1\linewidth]{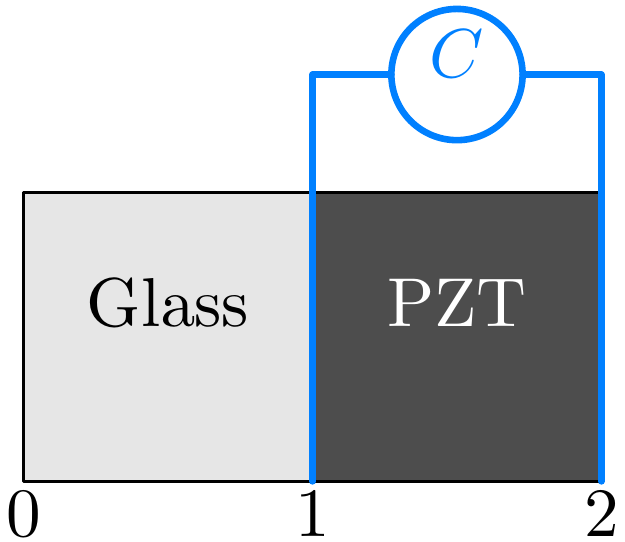}}
\end{minipage}
\hfill
\begin{minipage}[h]{0.75\linewidth}
\center{\includegraphics[width=1\linewidth]{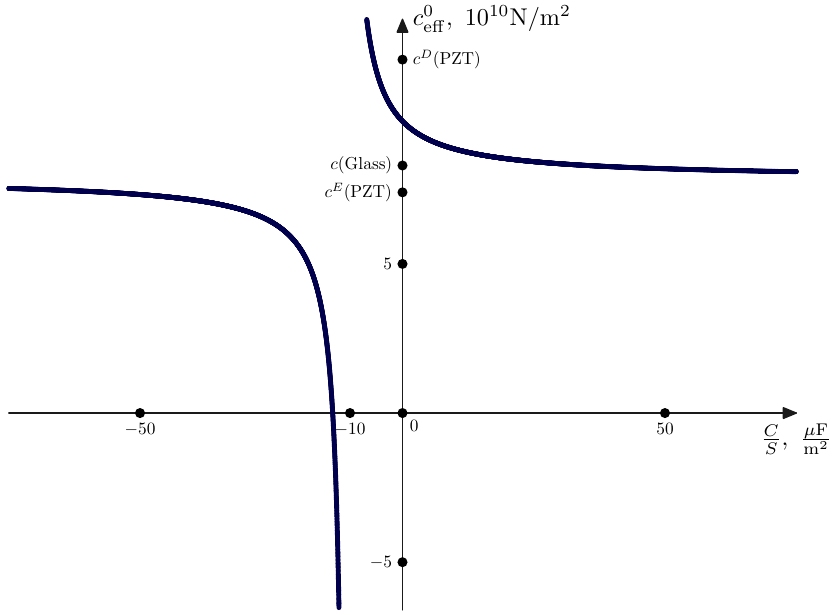}}
\end{minipage}
 \caption{Dependence of the quasistatic effective elastic constant $c_{%
\mathrm{eff}}^{0}$ on the external capacity $C$ for the glass / PZT
periodically bilayered structure where each PZT layer is electroded
and connected.} \label{fig1}
\end{figure}

Following \cite{1}, consider an example of the glass/PZT
periodically bilayered structure where each PZT layer is electroded
and connected as in the inset to Fig. \ref{fig1}. Let
$d_{1}=d_{2}=1$ mm. The material constants are given in \cite{1}.
The dependence (\ref{a9}) of the quasistatic effective elastic constant $c_{%
\mathrm{eff}}^{0}$ on $C$ is plotted in Fig.\ \ref{fig1}. It is seen
that $c_{\mathrm{eff}}^{0}$ is negative for
$\frac{C}{S}\in(\frac{C_{\infty}}{S}, \frac{C_{0}}{S}) $ where
\[
 \frac{C_{\infty}}{S}=-\frac{e^2d_1}{(c_1^{E}d_2+c_2^{E}d_1)d_2}-\frac{\varepsilon
 }{d_2},\ \ \ \
 \frac{C_{0}}{S}=-\frac{e^2}{d_2c_2^{E}}-\frac{\varepsilon}{d_2}\lb{zinf}
\]
are pole and zero of $c_{\mathrm{eff}}^{0}( \frac{C}{S})$. This
leads to a quasistatic absolute stopband which starts from $\omega
=0$ and extends to a certain nonzero $\omega$.

Let us examine in detail the evolution of the dispersion spectrum
with the decrease of $C<0$. Figure 2 displays a set of dispersion
spectra at different fixed values of $C<0$ which are compared with
the spectrum at $C=0$. It is seen that the values
$\frac{C}{S}\in(\frac{C_{\infty }}{S},0) $ first affect the second
branch $\omega(K)$ which changes the slope from negative to
positive. As $\frac{C}{S}$ approaches $ \frac{C_{\infty }}{S}\approx
10.67$ $\frac{ {\mu} \rm F}{\rm m^{2}}$, the group velocity $d\omega
/dK$ of the first branch at the origin point $\omega=0,~K=0$ tends
to infinity (along with the quasistatic phase velocity (\ref{es})).
This implies stiffening effect which can be observed in \cite{1}. In
the interval $\frac{C}{S}\in( \frac{C_{\infty
}}{S},\frac{C_{0}}{S})$, where $c_{\mathrm{eff}}^{0}<0$ (see Fig.
\ref{fig1}), there is the absolute stopband starting from the
quasistatic limit $\omega =0$. At certain
$\frac{C}{S}\in(\frac{C_{\infty }}{S}, \frac{C_{0}}{S})$, the first
branch which starts from $\omega\neq0$ becomes flat and then changes
slope from positive to negative, see Fig. 2 at $\frac{C}{S}=-11$
$\frac{ {\mu} \rm F}{\rm m^{2}}$ and $\frac{C}{S}=-12$ $\frac{ {\mu}
\rm F}{\rm m^{2}}$. For $\frac{C}{S}<\frac{C_{0}}{S}\approx 13.3$
$\frac{ {\mu} \rm F}{\rm m^{2}}$ the quasistatic elastic constant
$c_{\mathrm{eff}}^{0}$ and hence the quasistatic phase velocity
begin to grow from zero, so that the first dispersion branch
reappears at its quasistatic origin $\omega =0,~K=0$. Further
decrease of $C$ affects the position of the first stopband, as it is
observed in Fig. 2 for $\frac{C}{S}=-14$ $\frac{ {\mu} \rm F}{\rm
m^{2}}$, ..., $-40$ $\frac{ {\mu} \rm F}{\rm m^{2}}$.

\begin{figure}[h]
\begin{minipage}[h]{1\linewidth}
\begin{minipage}[h]{0.49\linewidth}
\center{\includegraphics[width=1\linewidth]{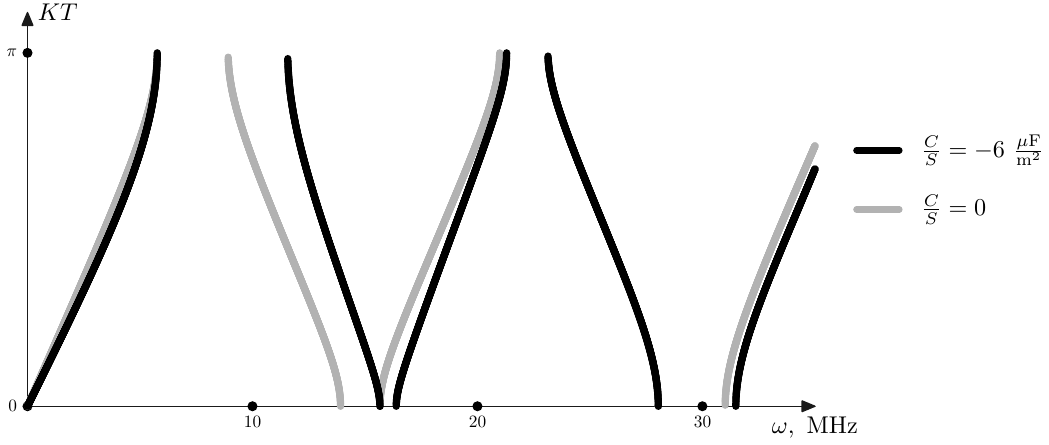}}
\end{minipage} \hfill
\begin{minipage}[h]{0.49\linewidth}
\center{\includegraphics[width=1\linewidth]{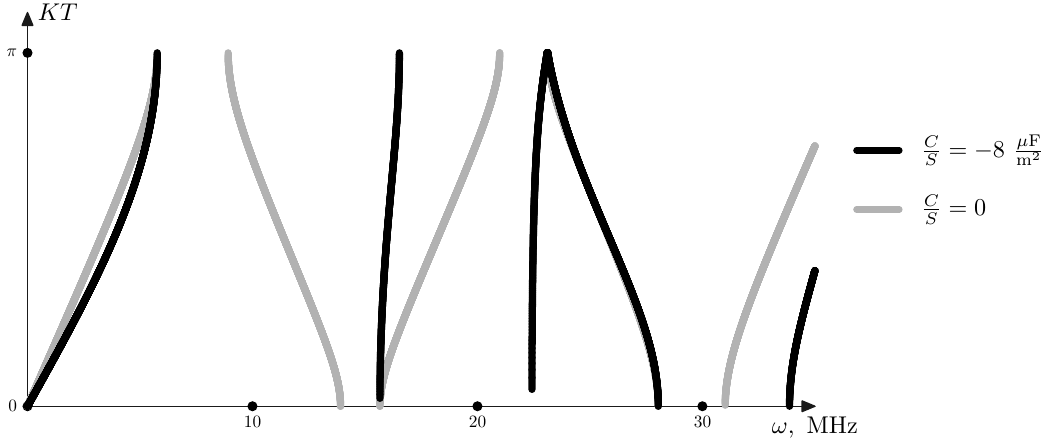}}
\end{minipage}
\end{minipage}
\vfil
\begin{minipage}[h]{1\linewidth}
\begin{minipage}[h]{0.49\linewidth}
\center{\includegraphics[width=1\linewidth]{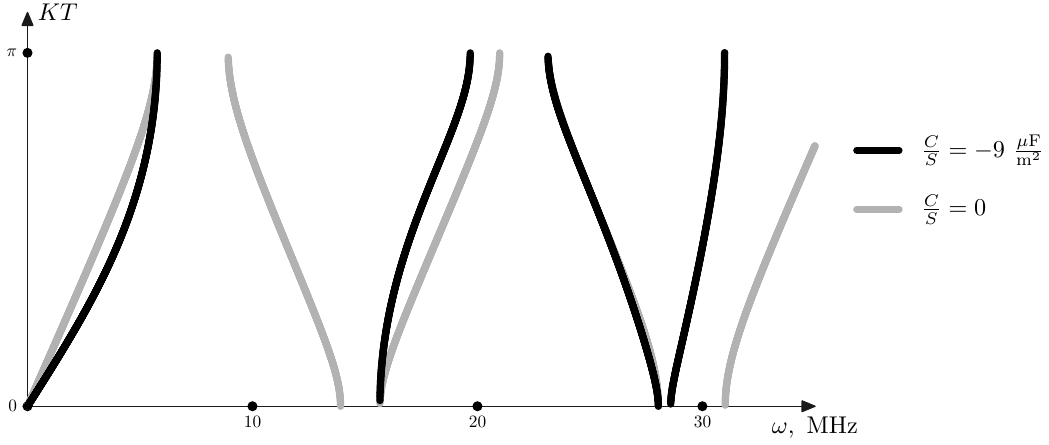}}
\end{minipage} \hfill
\begin{minipage}[h]{0.49\linewidth}
\center{\includegraphics[width=1\linewidth]{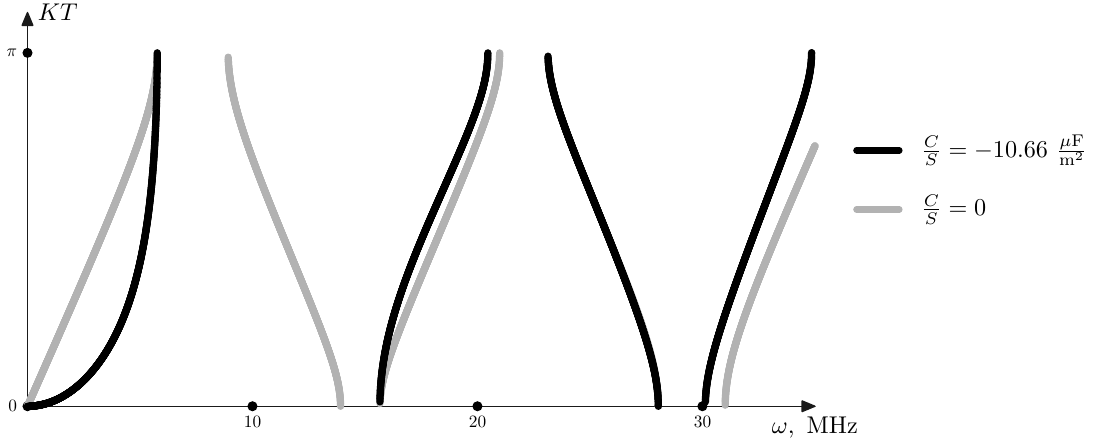}}
\end{minipage}
\end{minipage}
\vfil
\begin{minipage}[h]{1\linewidth}
\begin{minipage}[h]{0.49\linewidth}
\center{\includegraphics[width=1\linewidth]{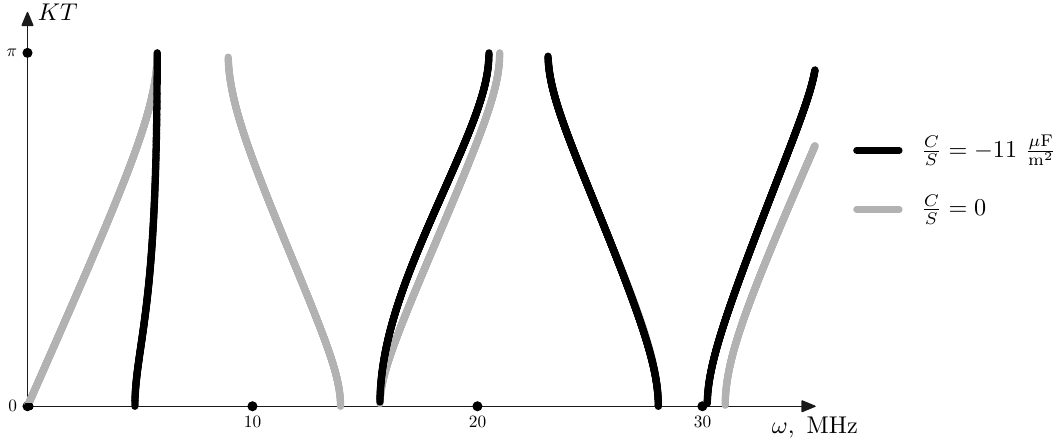}}
\end{minipage} \hfill
\begin{minipage}[h]{0.49\linewidth}
\center{\includegraphics[width=1\linewidth]{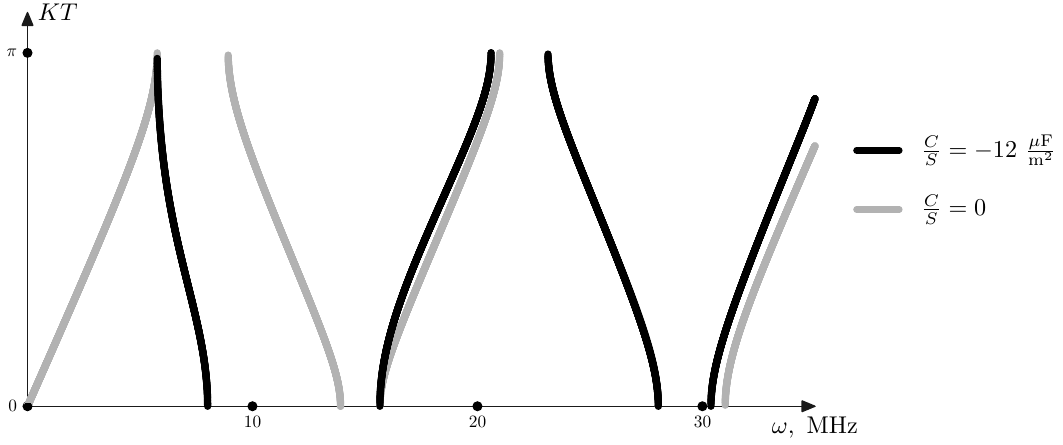}}
\end{minipage}
\end{minipage}
\vfil
\begin{minipage}[h]{1\linewidth}
\begin{minipage}[h]{0.49\linewidth}
\center{\includegraphics[width=1\linewidth]{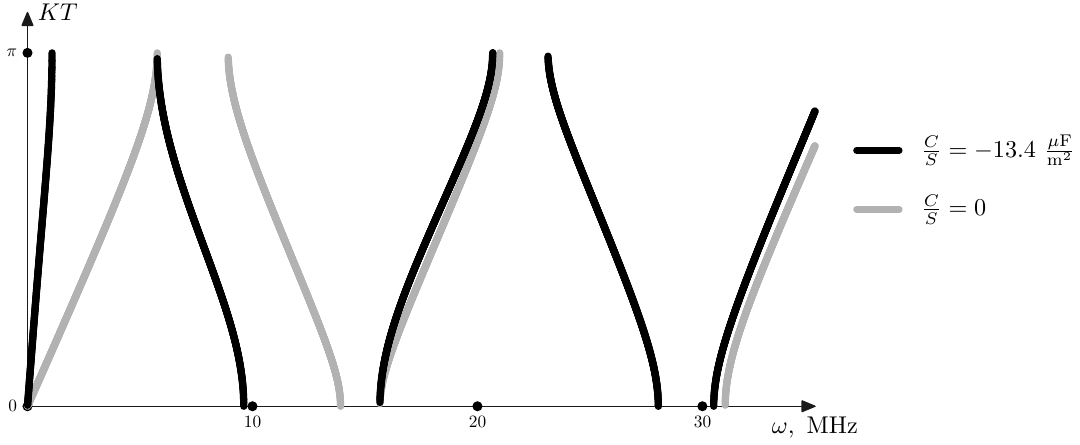}}
\end{minipage} \hfill
\begin{minipage}[h]{0.49\linewidth}
\center{\includegraphics[width=1\linewidth]{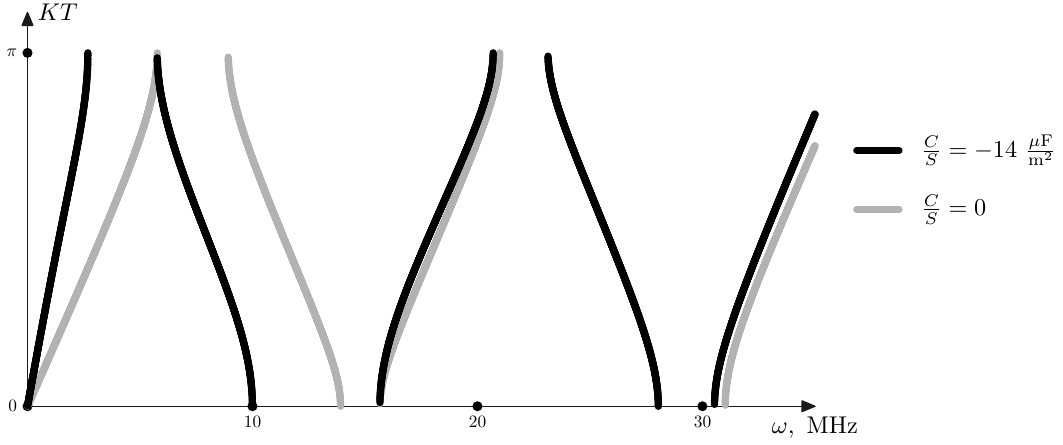}}
\end{minipage}
\end{minipage}
\vfil
\begin{minipage}[h]{1\linewidth}
\begin{minipage}[h]{0.49\linewidth}
\center{\includegraphics[width=1\linewidth]{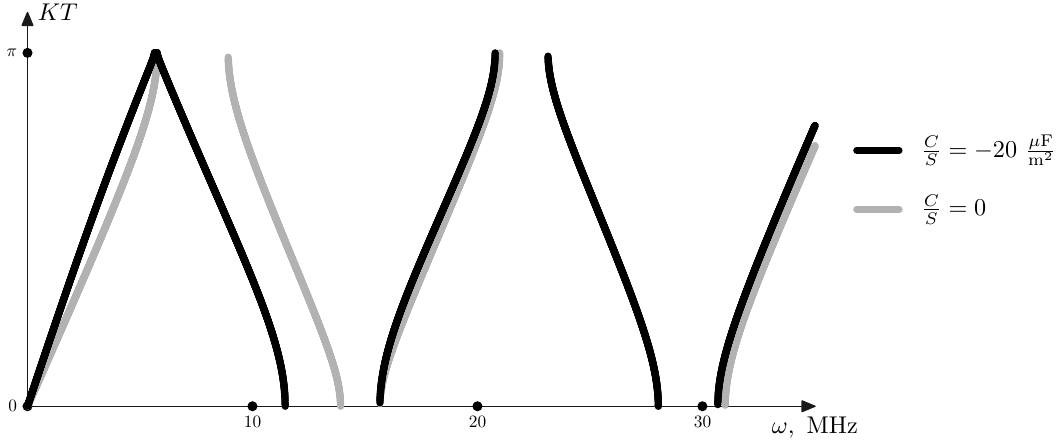}}
\end{minipage} \hfill
\begin{minipage}[h]{0.49\linewidth}
\center{\includegraphics[width=1\linewidth]{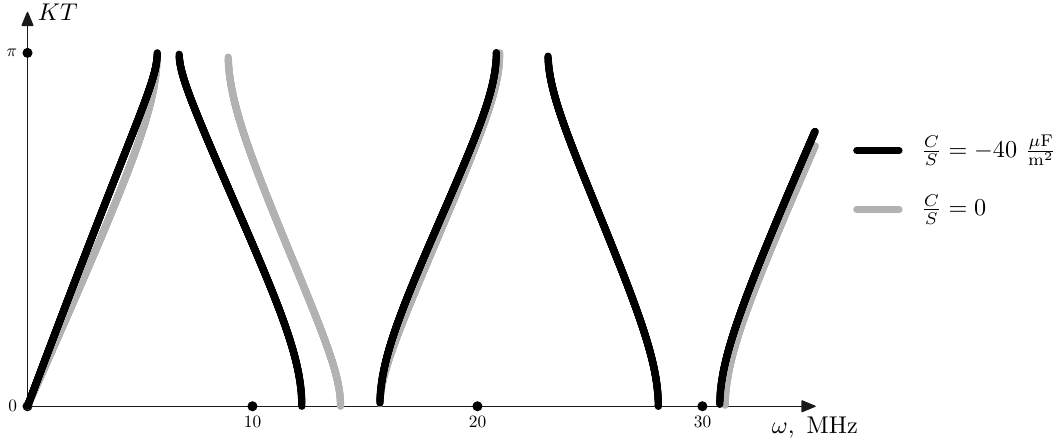}}
\end{minipage}
\end{minipage}
 \caption{Dispersion curves for the glass/PZT periodically bilayered structure (see inset in Fig. \ref{fig1})
 for different negative capacities $C$ (black lines) in comparison with the spectrum for $C=0$ (grey lines).} \label{fig1_1}
\end{figure}

\section{Conclusions}

A possibility has been shown of the quasistatic absolute stopband,
the infinite group velocity and the flat bands in the Floquet-Bloch
spectrum of a periodic structure of elastic and piezoelectric layers
where the faces of each piezoelectric layer are electroded and
connected via a circuit with negative capacity $C$.

\begin{acknowledgments}
This work has been conducted in the framework of the project MIRAGES
ANR-12-BS09-0015 with the support of the competitiveness cluster
Aerospace Valley. The authors grateful to A.-C.~Hladky and B. Dubus
for the useful discussions.
\end{acknowledgments}


\end{document}